# POSSIBLE ROUTES FOR THE FORMATION OF PREBIOTIC MOLECULES IN THE HORSEHEAD NEBULA

## *POSSÍVEIS ROTAS PARA A FORMAÇÃO DE MOLÉCULAS PREBIÓTICAS NA NEBULOSA CABEÇA DE CAVALO*

LUCIENE DA SILVA COELHO[1]; AMÂNCIO CÉSAR DOS SANTOS FRIAÇA[2]; EDGAR MENDOZA[2,3]
1 – PLANETÁRIO JUAN BERNARDINO MARQUES BARRIO, INSTITUTO DE ESTUDOS SOCIOAMBIENTAIS, UNIVERSIDADE FEDERAL DE GOIÁS; 2 – INSTITUTO DE ASTRONOMIA, GEOFÍSICA E CIÊNCIAS ATMOSFÉRICAS, UNIVERSIDADE DE SÃO PAULO; 3 – OBSERVATÓRIO DO VALONGO, UNIVERSIDADE FEDERAL DO RIO DE JANEIRO
*lucienecoelho@ufg.br; amancio.friaca@iag.usp.br; emendoza@astro.ufrj.br*

*Abstract - This article presents the results of a theoretical study concerning interstellar molecules which are useful for the bookkeeping of the organic content of the universe and for providing a glimpse into prebiotic conditions on Earth and in other environments in the universe. Performing the PDR Meudon code, we explored production channels for astrobiological relevant nitrogen-bearing cyclic molecules (N-heterocycles), e. g. pyrrole and pyridine. The present simulations demonstrate how the exploration of a few possible routes of production of N-heterocycles resulted in significant abundances for these species. One particularly efficient class of channels for the production of N-heterocycles incorporates polycyclic aromatic hydrocarbons (PAHs) as catalysts. Thereby, an exploration of a variety of production paths should reveal more species to be target of astrophysical observations.*

*Keywords: Astrochemistry and Astrobiology: Cosmic Prebiotic Chemistry. Organic Molecules. ISM: PDRs. Simulations.*

*Resumo – Este trabalho apresenta o resultado de um estudo teórico sobre moléculas interestelares que são úteis para a contabilidade do conteúdo orgânico do universo e para fornecer um vislumbre das condições pré-bióticas na Terra e em outros ambientes do além do nosso planeta. Utilizando o código PDR Meudon, nós exploramos rotas de formação de moléculas heterocíclicas nitrogenadas, como o pirrol e a piridina, as quais possuem potencial astrobiológico. As simulações demonstraram como a exploração de poucas rotas possíveis de produção de moléculas heterocíclicas nitrogenados resulta em abundâncias significantes destas espécies. Rotas particularmente efetivas para a produção de heterocíclicos nitrogenados envolvem os hidrocarbonetos policíclicos aromáticos como catalisadores. Assim, uma varredura por diversas rotas de produção deve revelar mais espécies a serem alvo de observações astrofísicas.*

*Palavras-chave: Astroquímica e Astrobiologia: Química Prebiótica Cósmica. Moléculas Orgânicas. ISM: PDRs. Simulações.*

## I. INTRODUCTION

One of the minimal traits of any living system is the presence of some mechanism for information storage, reading and self-replication. In terrestrial life, nucleic acids (DNA and RNA) are part of the machinery responsible for the informational function already mentioned (SECKBACH *et al.* 2000). It is not surprising that the nitrogen atom participates in the composition of molecules that store information. Since nitrogen, in contrast to carbon and oxygen, has an odd number of valence electrons, and introduces asymmetries in the molecular structure of any carbon compound. The existence of asymmetries is a necessary condition for information storage since a structure formed by the repetition of identical sub-units would not allow the writing of instructional sequences (NELSON & COX, 2012).

The cosmic abundance of the N atom is high enough to be found it in a huge variety of molecular forms, from the simplest species like CN (ADAMS, 1941); formamide, $NH_2CHO$ (LÓPEZ-SEPULCRE *et al.* 2019) to most complex structures, such as the polycyclic aromatic nitrogen heterocycles (PANHs), as some of which have been analyzed in Murchison meteorite samples (GLAVIN & BADA, 2004). PANHs are compounds derived from the polycyclic aromatic hydrocarbons (PAHs) when N atoms are part of the aromatic rings. PAHs are widely spread in the ISM, from stellar atmospheres up to galaxies at high redshifts (SHIVAEI *et al.* 2017). They constitute an efficient form to accumulate carbon in solid phase across the Universe, and their abundance and ubiquity are partially due to the high stability against the dissociative effects of ionizing agents (ALLAMANDOLA *et al.* 1999; SALAMA *et al.* 1999; TIELENS, 2008, 2013). PAHs are so ubiquitous that most of the mid-infrared emission in galaxies with star formation is dominated by strong emission features generally attributed to PAH bands (SPOON *et al.* 2002; TIELENS *et al.* 2004; PEETERS *et al.* 2004).

PANHs are especially relevant for astrobiology and they are closely related to the purine ($C_5H_4N_4$) and pyrimidine ($C_4H_4N_2$) nucleobases, which are part of the DNA and RNA molecular structures. These species are probably formed from the polymerization of small molecules such as acetylene ($C_2H_2$), nitrogen atoms being incorporated via the substitution of acetylene by cyanic acid (HCN) (PRASAD & HUNTRESS, 1980; FRENKLACH & FEIGELSON, 1989; HUDGIN *et al.*, 2005; CHEN *et al.*, 2008). The computational study of Ricca *et al.* (2001) has shown that minor units of PANHs could be synthesized via successive reactions of $C_2H_2$ and HCN under the stimulus of UV photons or cosmic rays. Laboratory experiments have also demonstrated that UV photoirradiation of icy samples, containing water ($H_2O$), ammonia ($NH_3$), and benzene ($C_6H_6$) or naphthalene ($C_{10}H_8$), leads to the formation of N and O-heterocycles, such as pyridine ($C_5H_5N$) and phthalide ($C_8H_6O_2$) (MATERESE *et al.*, 2015). In addition, PAHs also have a potential to synthetize amino acids. Chen *et al.* (2008) identified 13 different amino acids produced from the irradiation of ices containing $C_{10}H_8$, $H_2O$ and $NH_3$.

The spectroscopy of PANHs requires high sensitivity to differentiate their spectral signatures from those corresponding to PAHs. Hudgins *et al.* (2005) reported a shift in the peak position of PAHs at 6.2 μm, mainly due to typical signatures of PANHs, when a C atom is replaced by N. Laboratory measures have elucidated the different vibrational modes and spectroscopy of PANHs, whose peak positions are ranged between 6.7 – 10 micron (MATTIODA *et al.* 2005, 2017). The identification of a peak shift in that band, towards sources as the Horsehead Nebula, might reveal the presence, in the interstellar medium, of PANHs, which have already been found in meteorites (STOKS & SCHWARTZ, 1981) and suggested to be present in starburst-dominated galaxies (CANELO *et al.*, 2018).

If the substituted N is located on the periphery of the PAH, the resulting PANH is quite reactive, which may lead to the production of other N-heterocycles, as pyrroles, pyridines, pyrimidines or purines (e. g. MATERESE *et al.* 2015). Going back further than the RNA world, there could be a PANH world, or an aromatic world (EHRENFREUND *et al.* 2006). In this pre/protobiotic scenario, assembling PAH/PANH rich material could perform the transition from nonliving to living matter. For instance, PANHs may constitute a precursor to biological information carriers.

So far, the search for N-heterocycles has not yielded a conclusive identification in the gas phase of the interstellar medium (ISM), despite the multiple surveys performed in the sub-mm domain of the electromagnetic spectrum (SIMON & SIMON, 1973).

Kuan *et al.* (2003) searched for pyrimidine in three massive star-forming regions: Sgr B2(N), Orion KL and W51 e1/e2, however from the weak emission detected by those authors, only upper limits were derived for pyrimidine. Charnley *et al.* (2005) conducted research on pyridine ($C_5H_5N$), quinoline ($C_9H_7N$) and isoquinoline ($C_9H_7N$) towards circumstellar envelopes of carbon-rich stars; however, as in the research carried out by Kuan *et al.* (2003), only upper limits were obtained for these species. Experiments of UV irradiation reveal an important property on the lifetime of N-heterocycles. Peeters *et al.* (2004) found that gas samples of these molecules are rapidly destroyed with an increasing number of N atoms in the ring.

Astrochemical models based on experimental photodissociation rates indicates that complex organic molecules (COMs) should exhibit a short half-life in protoplanetary disk; for instance, in disk regions at ~ 10 AU and 70 AU from the protostar, $CH_3OH$ molecules survive 6.2 yr and 307.9 yr, respectively (ANDRADE *et al.* 2010). Interferometric observations of protoplanetary disks have revealed that various of them have rings and gaps; Leemker *et al.* 2021 discuss the implications that $HCO^+$ and $H^{13}CO^+$, among others molecules like $CH_3OH$ and $CH_3CHO$, would have on the snowline of protoplanetary disks, which is the inner boundary region where molecules condensate on dust grains forming icy mantles (ZHANG & JIN, 2015). Thus, observational and semi-empirical studies reveal important aspects on the presence and distribution of molecules in disks.

This work aims to conduct a study about the possibility of the formation of prebiotic N-heterocycles and other COMS in the interstellar medium, specifically, in the Horsehead Nebula, considering the physical conditions of molecular clouds and photodissociation regions (PDRs).

## II. METHOD

We used *the PDR Meudon code* to calculate the UV-driven chemistry of interstellar clouds considering the physical and chemical conditions of environments such as PDRs. The *PDR Meudon code* allows simulating regions as stationary plane-parallel slabs of gas and dust illuminated by radiation fields, which can be the Interstellar Standard Radiation Field. Heating (Photoelectric effect on grains, cosmic rays) and cooling (infrared and millimeter emission) processes contribute to the thermal balance. The output of *the PDR Meudon* includes gas properties like temperature, ionized fraction, chemical abundances and column densities (e.g., LE BOURLOT *et al.*, 1993; LE PETIT *et al.*, 2006; GONZALEZ GARCIA *et al.*, 2008).

The Horsehead Nebula was chosen as the benchmark for this study because of its relatively simple physics and geometry. It is a PDR, with some regions having a nearly plane-parallel geometry. We have assumed a fixed temperature of 15 K and parameters representative of the Horsehead Nebula, i.e., $G = 60\ G_0$, where $G_0$ is rather the flux of ISM as measured by Habing (1968), $A_V = 10$ mag to the cloud center, and a total hydrogen density of $10^4$ cm$^{-3}$. In addition, the classic parameterized modeling by Cordiner *et al.* (2007) for this object, well known for its rich atomic, molecular and dust emission observed from visible to sub-mm wavelength (GERIN *et al.*, 2009), make this nebula an ideal target to test any model for the formation of the molecular lines. Likewise, it is possible to extend the studies on the region, based on the results obtained by Goicoechea *et al.* (2009) and Gerin *et al.* (2009) considering this region as an archetype of molecular clouds. As a whole, the Horsehead Nebula is a complex source that harbors various environments, so that it can be divided into three distinct parts for the study of its chemistry, a nucleus, a PDR region and a PAH region, as modeled by Le Gal *et al* (2017), using *the PDR Meudon code*. This configuration favors the formation of COMs in the innermost region of the cloud, as it is a region more protected from the incidence of far-ultraviolet radiation.

In this work, we used all known reactions of formation and destruction for each specie along with the chemical precursors, databases such as UMIST and KIDA (MELROY *et al.* 2013, WAKELAM *et al.* 2015). The entire chemical network used here has 5238 reactions, 318 species and 14 elements.

## III. RESULTS

Some crucial characteristics make the ISM very promising for the emergence of complex chemistry. First of all, there is plenty of ultraviolet (UV) radiation, cosmic rays and shocks to provide the energy necessary for endothermic reactions. As a consequence, the interstellar chemistry is rich in species that require high energies for their formation, not only ions but also radicals, e.g., methylidyne (CH), methylene ($CH_2$), hydroxyl (OH) and cyanogen (CN).

*3.1 – Nitriles*

As we were interested in the synthesis of N-heterocycles, we explored production channels for pyrrole ($C_4H_5N$) and pyridine ($C_5H_5N$), the simplest N-heterocycles with 4 and 5 Carbon atoms, respectively. In particular, we consider the formation of pyridine from pyrrole via the reaction $CH + C_4H_5N \rightarrow C_5H_5N + H$ as suggested by Soorkia *et al.* (2010). In the first place, we studied the formation of pyrrole. One of the routes to produce it involves 2-butenal (or crotonaldehyde, $CH_3CHCHCHO$) as a precursor, which is a plausible route since aldehydes molecules have been already identified in the ISM. For instance, in the star-forming region Sagittarius B2(N), Hollis *et al.* (2004) reported the detection of propenal ($CH_2CHCHO$) and propanal ($CH_3CH_2CHO$), with similar structures and functional groups to the 2-butenal molecule. However, the production of pyrrole from 2-butenal resulted in very low abundances with $10^{-22}$ $C_4H_5N$ particles relative $H_2$ in the cloud core.

Another precursor to produce pyrrole is s-triazine, also called 1,3,5-triazine ($C_3H_3N_3$), the production process begins much more efficient when presents both reaction from 2-butenal and from s-triazine than the production only from 2-butanal, as we can see in Figure 1. However, the resulting abundance of pyrrole is too low with $10^{-18}$ $C_4H_5N$ particles relative hydrogen in the cloud core.

Figure 1 - Horsehead Nebula PDR model. Abundance relative to $H_2$ for pyrrole production from 2-butenal - blue line - and from s-triazine plus the 2-butenal - green line - paths of production as a function of extinction in V band, $A_V$, where 1 refers to the core of the cloud and -3,5 to the edge

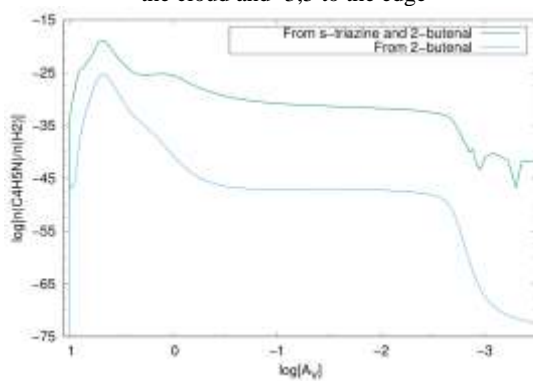

Source: Authors, 2021.

We also studied the pyridine formation from pyrrole, considering the route for pyrrole production which includes s-triazine as a precursor; however, the resulting abundances of pyridine are extremely low ($10^{-25}$ $C_5H_5N$ particles relative to $H_2$) in the cloud core, as shown in Figure 2. As we can see, also the pyrrole abundances keep being low.

It is interesting to note that the inclusion of a new species (pyridine, in this case) in the chemical reactions network could change the abundances of other molecules, even if these molecules are direct precursors (pyrrole, here) to the additional species. The new molecule participates in a network of reactions, some of which result in simpler species, which end up helping in the formation of more complex ones, including the direct precursors of the new species. This fact explains the differences of the relative abundances of pyrrole in Figure 1 (pyridine not included in the chemical network) and in Figure 2 (pyridine included in the chemical network).

Figure 2 - Horsehead Nebula PDR model. Abundance relative to $H_2$ for pyrrole and pyridine production as a function of extinction in V band, $A_V$, where 1 refers to the core of the cloud and -3,5 to the edge

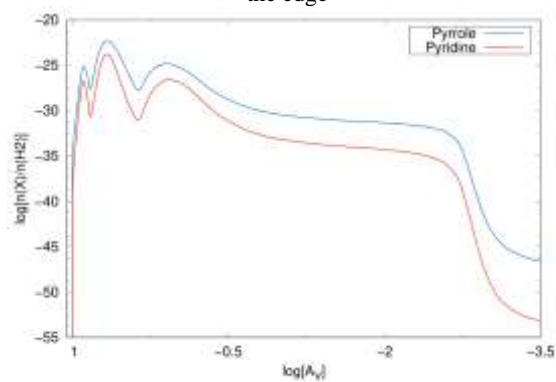

Source: Authors, 2021.

*3.2 – The role of PAHs*

The fact that PAHs are abundant molecules allowing the substitution of the C and H atoms by other radicals and atoms make them important precursors of PANHs and N-heterocycles.

The emission from 6.2 μm PAHs is probably dominated by species with molecular sizes, given by the number of C atoms, between 60 and 90 carbon atoms (Hudgins *et al.* 2005). Therefore, following the suggestion by Flower *et al.* (2003), we adopted circumcoronene, $C_{54}H_{18}$, as a representative molecule of interstellar PAHs. We considered a carbon fraction in PAHs of 15% in the interstellar medium as a compromise with estimates from other studies, e.g., 30-50% (PUGET & LEGER, 1989), 3-15% (ALLAMANDOLA *et al.* 1989), and 20-30% (JOBLIN *et al.*, 1992). Using for carbon the solar (photospheric) abundance $N_C/N_H=2.692 \times 10^{-4}$ (ASPLUND *et al.* 2009), under the simplification that all PAHs are $C_{54}H_{18}$, the corresponding PAH abundance relative to $H_2$ is $1,496 \times 10^{-6}$. Therefore, we take $10^{-6}$ particles relative to $H_2$ as the initial PAH abundance in our models.

PAHs are molecules that can resist interstellar radiation thanks to the bond energy of carbon atoms (8 eV) and the molecular backbone produced by them. In the case of absorption of UV photons, PAHs can rearrange their structures to form more stable structures. The aromatic skeleton plays a role in redistributing the incident energy of photons through different vibrational modes offering photostability against ionizing radiation (LEGER *et al.* 1989); for instance, a PAH with 50 carbon atoms could survive 1.1 Gyr in the Galactic radiation field (ALLAIN *et al.* 1996b). This is much longer than the lifetime for a typical nebula like the Horsehead Nebula. However, although the carbon skeleton of the PAH can be considered as a fixed structure, PAHs could suffer dehydrogenation, since the binding energy of hydrogen atoms is much lower, 4.5 eV (LEGER *et al.* 1989, MATTIODA *et al.* 2005), and PAHs may lose one or more hydrogen atoms.

After the dehydrogenation, it is possible to add a nitrogen atom in the PAH (RICCA *et al.* 2001, MATERESE *et al.* 2015), therefore generating a PANH. The addition of other radicals is also possible in a dehydrogenated PAH. Considering these mechanisms, we devised a chemical network involving PAHs (partially shown in Table 1), which leads to pyridine production. In Table 1, the binary reactions involving PAH and derived species have reaction rates of the form $k=k_0(T/300\ K)^{1/2}$. PAHs are not particularly reactive but their dehydrogenated forms or with incorporated atoms have kinetic reaction rates about a factor ten higher (LE PAGE *et al*., 2001; VUONG & FOING, 2000).

The results are displayed in Figure 3. We can see that PAHs are dehydrogenated throughout all the cloud and, also, PANHs have low abundances at the cloud edge and increase towards the center.

Table 1 – Reaction rates for chemistry involving circumcoronene and derived species

| Reagents | Products | $k_0$ (cm$^3$s$^{-1}$) |
|---|---|---|
| $C_{54}H_{18} + H$ | $C_{54}H_{17} + H_2$ | $5.00 \times 10^{-8}$ |
| $C_{54}H_{18} + C_2H$ | $C_{54}H_{17} + C_2H_2$ | $5.00 \times 10^{-8}$ |
| $C_{54}H_{17} + HCH$ | $C_{55}H_{18}N$ +photon | $5.00 \times 10^{-7}$ |
| $C_{55}H_{18}N + C_2H_2$ | $C_{57}H_{19}N + H$ | $5.00 \times 10^{-7}$ |
| $C_{57}H_{19}N + C_2H_4$ | $C_{54}H_{18} + C_5H_5N$ | $5.00 \times 10^{-7}$ |

Source: Authors, 2020.

Figure 3 - Horsehead Nebula PDR model. Abundance relative to $H_2$ for circumcoronene, its dehydrogenated forms as well as the PANHs formed from circumcoronene as a function of extinction in V band, Av, where 1 refers to the core of the cloud and -3,5 to the edge

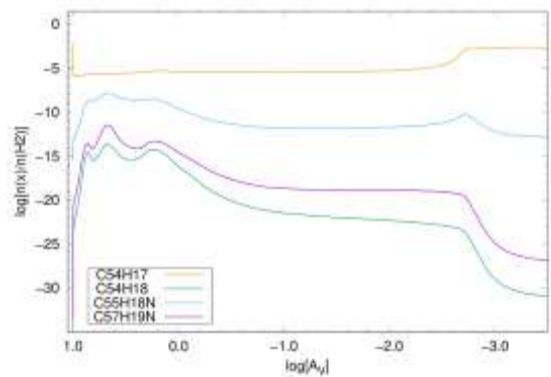

Source: Authors, 2021.

The PANH molecule with formula $C_{57}H_{19}N$ is not very abundant ($10^{-12}$ $C_{57}H_{19}N$ particles relative hydrogen in the cloud core), nevertheless it is twice more abundant than the hydrogenated form of the PAH ($C_{54}H_{18}$). Since $C_{57}H_{19}N$ is a precursor of pyridine, the introduction of a PAH represents a very effective pyridine channel production, at least in the cloud core. At the end of the reaction, when the pyridine is released from the PAH, $C_{54}H_{18}$ is recovered. The PAH has played the role of a catalyst, promoting the encounter of common species such as HCN and hydrocarbons to generate N-heterocyclics.

Figure 4 compares the production of pyridine through two channels: from PAHs and pyrrole ($C_4H_5N$) previously synthesized. As we can see, the pyridine production through PAHs is much more efficient and in the innermost parts of the cloud ($A_V > 5$) it is about ten orders of magnitude higher.

Table 2 lists the column density [cm$^{-2}$] of pyrrole and pyridine for different production channels at several optical depths of Horsehead Nebula; Pyrrole and Pyridine refer to its production without PAH. Pyrrole$^+$ and Pyridine$^+$ are data from their formation with circumcoronene include. Pyrrole* and Pyridine* are column densities for their formation including circumcoronene, coronene and ovalene, following the same network of reactions as circumcoronene.

Figure 4 - Horsehead Nebula PDR model. Abundance relative to $H_2$ estimated for two different channels of pyridine production: both from the network of reactions of PAHs and directly from pyrrole as a function of extinction in V band, Av, where 1 refers to the core of the cloud and -3,5 to the edge

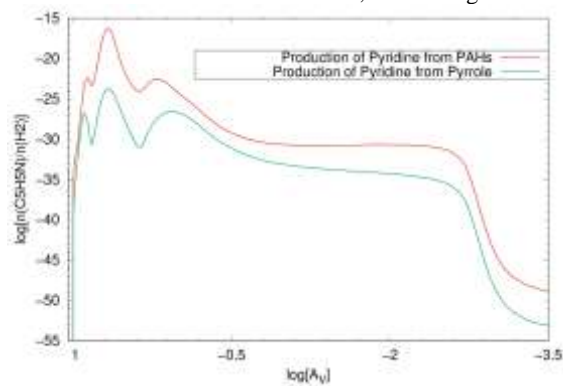

Source: Authors, 2021.

The results of our calculations indicate that PAHs can be an important intermediate species in producing nitrogen heterocycles, as illustrated by the high abundances obtained for pyridine in modeling the Horsehead Nebula. In the specific model, we considered, pyrrole dominates over pyridine in the outskirts of the cloud, but pyridine becomes more abundant in the inner regions, as seen from Figure 5. It is interesting that a similar behavior for the two species also appears in the photochemistry model for Titan's atmosphere by Krasnopolsky (2009, 2014).

Figure 5 - Horsehead Nebula PDR model. Abundance relative to $H_2$ comparing production of pyrrole and pyridine as a function of extinction in V band, AV, where 1 refers to the core of the cloud and -3,5 to the edge

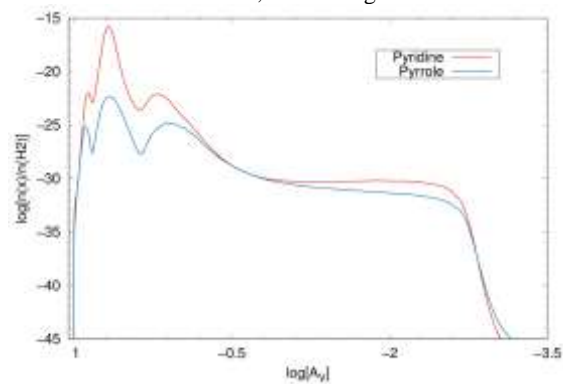

Source: Authors, 2021.

Table 2 – Column density [cm$^{−2}$] from pyrrole and pyridine, for different production channels at depths of Horsehead Nebula. *Pyrrole* and *Pyridine* refer to its production without PAH. Pyrrole$^+$ and Pyridine$^+$ are data from their formation with PAH circumcoronene include. Pyrrole* and Pyridine* are column densities for their formation including circumcoronene, coronene and ovalene

| Molecules | $A_V = 1$ mag | $A_V = 3$ mag | $A_V = 10$ mag |
|---|---|---|---|
| Pyrrole | $3.71 \times 10^{-4}$ | $6.17 \times 10^{-2}$ | $9.88 \times 10^7$ |
| Pyrrole$^+$ | $2.71 \times 10^{-3}$ | $1.58 \times 10^4$ | $8.52 \times 10^9$ |
| Pyrrole* | $2.85 \times 10^{-2}$ | $2.53 \times 10^5$ | $1.37 \times 10^{10}$ |
| Pyridine | $2.14 \times 10^{-10}$ | $2.51 \times 10^{-9}$ | $3.29 \times 10^{-5}$ |
| Pyridine$^+$ | $1.15 \times 10^{-4}$ | $2.01 \times 10^3$ | $1.14 \times 10^{12}$ |
| Pyridine* | $1.19 \times 10^{-3}$ | $1.20 \times 10^4$ | $1.50 \times 10^{13}$ |

Source: Authors, 2021.

As shown in Table 2, the column density of pyridine is about $10^{13}$ cm$^{-2}$ for the route of production with the PAHs pathway. The high abundance obtained for this molecule makes it a target for future observations. This value is close to the upper limits obtained for other types of objects – the ABG star IRC+10216 ($<7.3$ - $8.6 \times 10^{12}$ cm$^{-2}$) and the planetary nebula CRL 618 ($2.3$ - $2.7 \times 10^{13}$ cm$^{-2}$) – (CHARNLEY *et al.*, 2005). In view of the ubiquitous presence of PAHs and PANHs in interstellar and circumstellar environments, these species can catalyze the formation of heterocyclics with an incorporated nitrogen atom, but for two heterocyclics with two atoms, as pyrimidine, pyrimidone and uracil, the abundances are expected to be lower.

In general, several molecules that are the basic units of life are most easily synthesized in ISM than on Earth. Observational work should be allied with laboratory experiments and theoretical calculations in order to derive the spectrum in the infrared, millimeter and radio wavelengths, and to estimate the formation and survival of prebiotic molecules in the ISM, which eventually could be released in planet-like environments, providing the building blocks of life.

## IV. COMPARISON WITH OBSERVATIONS

To compare the results obtained in this study with both observations on Horsehead Nebula and other models already made for the same region, we used the results of Goicoechea *et al.* (2009) and Teyssier *et al.* (2003) for some with some smaller and easier-to-observe hydrocarbons. Comparisons were summarized in Table 3 and test the reliability of the data.

Table 3 – Column densities [cm$^{-2}$], and abundances [n(X)/n(H)] for some well-known molecules in PDR of the Horsehead Nebula, like CCH, c-C3H2, CS and HCS+, to compare measurements and models

| Autor | Data Type | CCH | *c*-C3H2 |
|---|---|---|---|
| Goicoechea, 2009 | Column Density | $(1.1 \pm 0.3) \times 10^{14}$ | $(9.5 \pm 5.0) \times 10^{12}$ |
|  | Abundance | $1.5 \times 10^{-8}$ | $1.3 \times 10^{-9}$ |
| Teyssier, 2003 | Measured | $(1.8 \pm 0.2) \times 10^{14}$ | $(1 – 2) \times 10^{13}$ |
|  | Model | $(1 – 2.5) \times 10^{14}$ | $(0.5 – 1.5) \times 10^{12}$ |
| Authors, 2020 | Column Density | $8.3 \times 10^{13}$ | $8.4 \times 10^{11}$ |
|  | Abundance | $2.6 \times 10^{-8}$ | $2.5 \times 10^{-8}$ |
|  |  | **CS** | **HCS+** |
| Goicoechea, 2009 | Column Density | $(1.12 \pm 1.0) \times 10^{14}$ | $(6.8 \pm 0.5) \times 10^{11}$ |
|  | Abundance | $3.9 \times 10^{-9}$ | $2.3 \times 10^{-11}$ |
| Authors, 2020 | Column Density | $6.9 \times 10^{14}$ | $1.5 \times 10^{11}$ |
|  | Abundance | $6.1 \times 10^{-8}$ | $1.8 \times 10^{-11}$ |

Source: Goicoechea *et al.* 2009, Teyssier *et al.* 2003 and Authors, 2021.

When we researched and compared the abundances derived from the observations of Goicoechea *et al.* (2009) and those obtained in this work, we have a satisfactory agreement with the data. There is mainly a concordance between Ethynyl radical (CCH) molecules and HCS+ whose abundances agree in units. In other abundances there is a minor difference, about ten-fold, calculated by the model.

Likewise, there is a discrepancy relation with column densities, but it is only one order of magnitude. If we compare the column densities obtained in this study and those obtained by the model of Teyssier *et al.* (2003) the data agreement is even better, smaller than ten magnitudes.

In both cases, the differences between the presented models, observations and this work are within the acceptable and we can say that the model built here is quite reliable.

## V. CONCLUSIONS

In this work, we developed a theorethical study about prebiotic molecules in the Horsehead nebula. Our goal was to investigate the abundance of molecules with a prebiotic role in environments like the Horsehead nebula, in view of issues such as the origin of life on Earth and other places in the Universe.

The Horsehead nebula is an archetype of photodissociation region (PDR), with physical and chemical parameters well-known in the literature. Thus, we could modeling this region using the PDR Meudon code which is widely used since can simulate PDRs regions accurately considering the physical and chemical properties of a given system.

At first, we add the in the PDR Meudon code all known reactions of formation and destruction of the pyrrole and pyridine, however, this model presented abundances relatives to H$_2$ too low to be detected through observations. So, we developed a net of reactions for polycyclic aromatic hydrocarbons (PAHs) turning them into catalysts for the formation of these prebiotic molecules. For this, we use the circumcoronene as a representative of long-life PAHs. Those net of reactions were made considering that PAHs are hardly destroyed but can be dehydrogenated.

The results of our calculations indicate that PAHs can be an important intermediate species in producing nitrogen heterocycles, as illustrated by the high abundances obtained for pyridine in modeling the Horsehead Nebula.

Our results were compare with simulations and observations reported by Goicoechea *et al.* (2009) and Teyssier *et al.* (2003) and we have a satisfactory agreement with the data, in which there was, at most, a difference of about ten-fold, calculated by the model.

Therefore, after verifying that the model agrees with the observational data for the region, the present simulations demonstrate how the exploration of a few possible routes of production of N-heterocycles resulted in significant abundances for these species. One particularly efficient class of channels for the production of N-heterocycles incorporates polycyclic aromatic hydrocarbons (PAHs) as catalysts. Thereby, an exploration of a variety of production paths should reveal more species to be target of astrophysical observations.

## VI. REFERENCES

ADAMS, W. S. Some Results with the COUDÉ Spectrograph of the Mount Wilson Observatory, **The Astrophysical Journal** 93, 11, 1941.

ALLAIN, T., LEACH, S., SEDLMAYR, E. Photodestruction of PAHs in the interstellar medium. II. Influence of the states of ionization and hydrogenation, **Astronomy & Astrophysics** 305, 616, 1996b.

ALLAMANDOLA, L.J., TIELENS, A.G.G.M., BARKER, J.R. Interstellar polycyclic aromatic hydrocarbons - The infrared emission bands, the excitation/emission mechanism, and the astrophysical implications, **The Astrophysical Journal Supplement Series**, 71, 733-775, 1989.

ALLAMANDOLA, L.J., HUDGINS, D.M. and SANDFORD, S.A. Modeling the Unidentified Infrared Emission with Combinations of Polycyclic Aromatic Hydrocarbons, **The Astrophysical Journal**, 511, L115, 1999.

ANDRADE, D.P.P., ROCCO, M.L.M., BOECHAT-ROBERTY, H. M. X-ray photodesorption from methanol ice, **Monthly Notices of the Royal Astronomical Society**, 409, 1289, 2010.

ASPLUND, M., GREVESSE, N., SAUVAL, A.J., SCOTT, P. The chemical composition of the Sun. **Annual Reviews of Astronomy and Astrophysics**, 47, 481-522, 2009.

CANELO, C. M.; FRIAÇA, A. C. S.; SALES, D. A.; PASTORIZA, M. G.; RUCHEL-DUTRA, D. Variations in the 6.2 μm emission profile in starburst-dominated galaxies: a signature of polycyclic aromatic nitrogen heterocycles (PANHs)? **Monthly Notices of the Royal Astronomical Society**, 475, 3746, 2018.

CHARNLEY, S.B., KUAN, Y.J., HUANG, H.C., BOTTA, O., BUTNER, H.M., COX, N., DESPOIS, D., EHRENFREUND, P., KISIEL, Z., LEE, Y.Y., MARKWICK, A.J., PEETERS, Z., RODGERS, S.D. Astronomical searches for nitrogen heterocycles, **Advances in Space Research** 36, 137, 2005.

CHEN, Y.J., NUEVO, M., YIH, T.S.; IP, W.H., FUNG, H.S., CHENG, C.Y., TSAI, H.R., WU, C.Y. Amino acids produced from the ultraviolet/extreme-ultraviolet irradiation of naphthalene in a H2O+NH3 ice mixture, **Monthly Notices of the Royal Astronomical Society** 384, 605, 2008.

CORDINER, M.A., MILLAR, T.J., HERBST, E., CHUIMIN, R.Ni., WALSH, C. **Molecular anion chemistry in interstellar and circumstellar environments**. Molecules in Space and Laboratory, meeting held in Paris, France, 2007.

EHRENFREUND, P., RASMUSSEN, S., CLEAVES, J., CHEN, L. Experimentally tracing the key steps in the origin of life: The aromatic wold, **Astrobiology** 6, 490, 2006.

FLOWER D.R., PINEAU DES FORÊTS, G. The influence of grains on the propagation and structure of C-type shock waves in interstellar molecular clouds, **Monthly Notices of the Royal Astronomical Society** 343, 390, 2003.

FRENKLACH, M., FEIGELSON, E.D. PAH formation in carbon-rich circumstellar envelopes, **The Astrophysical Journal** 341, 372, 1989.

GERIN, M., PETY, J., GOICOECHEA, J.R. The Horsehead Nebula, a template source for interstellar physics and chemistry, **Submillimeter Astrophysics and Technology:** a Symposium Honoring Thomas G. Phillips, edited by D. C. Lis, J. E. Vaillancourt, P. F. Goldsmith, T. A. Bell, N. Z. Scoville, & J. Zmuidzinas, vol. 417 of Astronomical Society of the Pacific Conference Series, pp. 165, 2009.

GLAVIN, D.P., BADA, J.L. Isolation of Purines and Pyrimidines from Murchison Meteorite Using Sublimation, **35th Lunar and Planetary Science Conference**, March 15-19, League City, Texas, abstract 1022, 2004.

GOICOECHEA, J.R., PETY, J., GERIN, E.A., HILY-BLANT, P., LE BOURLOT, J. The ionization fraction gradient across the Horsehead edge: an archetype for molecular clouds, **Astronomy & Astrophysics**, 498, 771, 2009.

GONZALEZ GARCIA, M., LE BOURLOT, J., LE PETIT, F., ROUEFF, E. Radiative transfer revisited for emission lines in photon dominated regions, **Astronomy & Astrophysics**, 485, 127, 2008.

HABING, H. J., The interstellar radiation density between 912Å and 2400Å, **Astronomical Institutes of The Netherlands**, 19, 421-431, 1968

HOLLIS, J.M., JEWELL, P.R., LOVAS, F.J., REMIJAN, A. Green Bank Telescope Observations of Interstellar Glycolaldehyde: Low-Temperature Sugar, **The Astrophysical Journal** 613, L45 2004.

HUDGINS, D.M., BAUSCHLICHER, C.W.Jr., ALLAMANDOLA, L.J. Variations in the peak position of the 6.2 μm interstellar emission feature: A tracer of N in interstellar polycyclic Aromatic Hydrocarbon Population, **The Astrophysical Journal** 632, 316, 2005.

JOBLIN, C., LEGER, A., MARTIN, P. Contribution of polycyclic aromatic hydrocarbon molecules to the interstellar extinction curve, **The Astrophysical Journal**, 393, L79, 1992.

KRASNOPOLSKY, V.A. A photochemical model of Titan`s atmosphere and ionosphere, **Icarus** 201, 226, 2009.

KRASNOPOLSKY, V.A. Chemical composition of Titan's atmosphere and ionosphere: Observations and the photochemical model, **Icarus** 236, 83, 2014.

KUAN, Y.J., YAN, C.H., CHARNLEY, S.B., KISIEL, Z., EHRENFREUND, P., HUANG, H.C. A search for interstellar pyrimidine, **Monthly Notices of the Royal Astronomical Society**, 345, 650, 2003.

LE GAL, R., HERBIST, E., DUFOUR, G., GRATIER, P., RUAUD, M., VIDAL, T.H.G., WAKELAM, V. A new study of the chemical structure of the Horsehead nebula: the influence of grains-surface chemistry, **Astronomy & Astrophysics**, 605, id.A88, 16 pp, 2017.

LE BOURLOT, J., PINEAU DES FORESTS, G., ROUEFF, E., FLOWER, D.R. Infrared and submillimetric emission lines from the envelopes of dark clouds, **Astronomy & Astrophysics**, 267, 233, 1993.

LE PAGE, V., SNOW, T. P., BIERBAUM, V. M. Hydrogenation and Charge States of PAHS in Diffuse Clouds. I. Development of a Model, **The Astrophysical Journal Supplement Series**, 132, 233, 2001.

LE PETIT, F., NEHMÉ, C., LE BOURLOT, J., ROUEFF, E. A model for Atomic and Molecular Interstellar Gas: The Meudon PDR Code, **Astrophysics Journal Supplements Series**, 164, 506, 2006.

LEGER, A., D'HENDECOURT, L., BOISSEL, P., DESERT, F.X. Photo-thermo-dissociation. I.-A general mechanism for destroying molecules, **Astronomy & Astrophysics**, 213, 351, 1989.

LEEMKER, M., VAN'T HOFF, M. L. R., TRAPMAN, L., VAN GELDER, M. L., HOGERHEIJDE, M. R., RUIZ-RODRIGUEZ D., VAN DISHOECK, E. F. Chemically tracing the water snowline in protoplanetary disks with HCO+, **Astronomy & Astrophysics**, 646, A3, 2021.

LÓPEZ-SEPULCRE, A., BALUCANI, N., CECCARELLI, C, CODELLA, C., DULIEU, F., THEULÉ, P. Interstellar Formamide (NH2CHO), a Key Prebiotic Precursor, **ACS Earth and Space Chemistry**. 3, 10, 2122-2137, 2019.

MATERESE, C.K., NUEVO, M., SANDFORD, S. A. N- and O-heterocycles Produced from the Irradiation of Benzene and Naphthalene in H20/NH3-containing Ices, **The Astrophysics Journal** 800, 116, 2015.

MATTIODA, A.L., HUDGINS, D.M., BAUSCHLICHER, C.W., ALLAMANDOLA, L.J. Infrared spectroscopy of matrix isolated polycyclic aromatic compounds and their ions. 7. Phenazine, a dual substituted polycyclic aromatic nitrogen heterocycle, **Advances in Space Research** 36, 156, 2005.

MATTIODA, A.L., BAUSCHLICHER, C.W., RICCA, A., BREGMAN, J., HUDGINS, D.M., ALLAMANDOLA, L.J. Infrared spectroscopy of matrix-isolated neutral polycyclic aromatic nitrogen heterocycles: The acridine series, **Spectrochimica Acta Part A** 181, 286, 2017.

McELROY, D., WALSH, C., MARKWICK, A.J., CORDINER, M.A., SMITH, K., MILLAR, T.J. The UMIST database for astrochemistry 2012. **Astronomy & Astrophysics**, vol 550, A36, 2013.

NELSON, D.L., COX, M.M. **Lehninger Principles of Biochemistry**, Publisher: W. H. Freeman, 2012.

PRASAD, S.S., HUNTRESS, W.T.Jr. A model for gas phase chemistry in interstellar clouds. II - Nonequilibrium effects and effects of temperature and activation energies, **Astrophysical Journal**, Part 1, vol. 239, p. 151-165, 1980.

PEETERS, E., SPOON, H.W.W., TIELENS, A.G.G.M. Polycyclic Aromatic Hydrocarbons as a Tracer of Star Formation?, **The Astrophysical Journal** 613, 986, 2004.

PUGET, J. L., LEGER, A. A new component of the interstellar matter - Small grains and large aromatic molecules, **Annual Review of Astronomy and Astrophysics**, 27, 161, 1989.

RICCA, A., BAUSCHLICHER, C.W., BAKES, E.L.O. A Computational Study of the Mechanisms for the Incorporation of a Nitrogen Atom into Polycyclic Aromatic Hydrocarbons in the Titan Haze, **Icarus** 154, 516, 2001.

SALAMA, F., GALAZUTDINOV, G.A., KRELOWSKI, J., ALLAMANDOLA, L.J., MUSAEV F. A. Polycyclic Aromatic Hydrocarbons and the Diffuse Interstellar Bands: A Survey, **The Astrophysics Journal** 526, 265, 1999.

SECKBACH, J., WESTALL, F., CHELA-FLORES, J. Introduction to Astrobiology: Origin, Evolution, Distribution and Destiny of Life in the Universe; **Kluwer Academic Publishers**, Dordrecht, The Netherlands, 2000. Introduction to Astrobiology.

SHIVAEI, I., REDDY, N.A., SHAPLEY, A.E., SIANA, B., KRIEK, M., MOBASHER, B., COIL, A.L., FREEMAN, W.R., SANDERS, R.L., PRICE, S.H., AZADI, M., ZICK, T. The MOSDEF Survey: Metallicity Dependence of PAH Emission at High Redshift and Implications for 24 μm Inferred IR Luminosities and Star Formation Rates at z~2, **The Astrophysics Journal** 837, 157, 2017.

SIMON, M.N., SIMON, M. Search for Interstellar Acrylonitrile, Pyrimidine and Pyridine, **The Astrophysics Journal** 184, 757, 1973.

SOORKIA, S., TAATJES, C.A., OSBORN, D.L., SELBY, T.M., TREVITT, A.J., WILSON, K.R., LEONE, S.R. Direct detection of pyridine formation by the reaction of CH (CD) with pyrrole: a ring expansion reaction, **Physical Chemistry Chemical Physics** 12, 31, 8750, 2010.

SPOON, H.W.W., KEANE, J.V., TIELENS, A.G.G.M., LUTZ, D., MOORWOOD, A.F.M., LAURENT, O. Ice features in the mid-IR spectra of galactic nuclei, **Astronomy & Astrophysics** 385, 1022, 2002.

STOKS, P.G., SCHWARTZ, A.W. Nitrogen-heterocyclic compounds in meteorites: significance and mechanisms of formation, **Geochimica et Cosmochimica Acta**, 45, 563, 1981.

TEYSSIER, D., FOSSÉ, D., GERINn, M., PETY, J., ABERGEL, A., HABART, E. Connection Between PAHs and Small Hydrocarbons in the Horsehead Nebula PhotoDissociation Region, in **SFChem 2002:** Chemistry as a Diagnostic of Star Formation, edited by C. L. Curry and M. Fich, p. 422, 2003.

TIELENS, A.G.G.M., PEETERS, E., BAKES, E.L.O., SPOON, H.W.W., HONY, S. PAHs and Star Formation, in Star Formation in the Interstellar Medium: In Honor of David Hollenbach, edited by D. Johnstone, F. C. Adams, D. N. C. Lin, D. A. Neufeeld, and E. C. Ostriker (2004), vol. 323 of **Astronomical Society of the Pacific Conference Series**, p. 135, 2004.

TIELENS, A.G.G.M. Interstellar polycyclic aromatic hydrocarbons molecules, **Annual Review of Astronomy and Astrophysics** 46, 289, 2008.

TIELENS, A.G.G.M. Interstellar PAHs and Dust, Planets, Stars and Stellar Systems Vol.5., **Springer Science**, 2013.

VUONG, M. H., FOING, B. H. Dehydrogenation of polycyclic aromatic hydrocarbons in the diffuse interstellar medium, **Astronomy and Astrophysics**, 363, L5, 2000.

WAKELAM, V., *et al*. The 2014 KIDA Network for Interstellar Chemistry. **The Astrophysical Journal Souplement Series**, vol 217, 20, 2015.

ZHANG, Y., & JIN, L. The evolution of the snow line in a protoplanetary disk. **The Astrophysical Journal** 802, 58, 2015.

VII. COPYRIGHT

The authors are the only ones responsible for the material included in this article.